\documentclass[12pt,german]{article}
\usepackage[pdftex]{graphics}
\usepackage{amsmath}
\usepackage[latin1]{inputenc}
\usepackage[small]{caption}
\usepackage{graphicx}
\begin{document}
\begin{center}
\Large{\bf Test of the Law of Gravitation at small Accelerations.} \\ 

\normalsize
\vspace{3.0cm}

H.Meyer (Bergische Universit\"at Wuppertal), E.Lohrmann, S.Schubert 
(Universit\"at Hamburg), W.Bartel, A.Glazov, B.L\"ohr, C.Niebuhr, E. W\"unsch 
(DESY), L.J\"onsson (University of Lund),
G. Kempf (Hamburgische Schiffbau-Versuchsanstalt). \\

\vspace{1.0cm}
\end{center}

\begin{center}
{\bf Abstract \\}
\end{center}
Newton's Law of Gravitation has been tested at small values $a$ of
the acceleration, down to $a \, \approx \, 10^{-10} \, m \, s^{-2}$,
 the approximate value of MOND's constant $a_0$. No deviations were found. \\

\section{Introduction}
The nature of Dark Matter is one of the central questions in astrophysics at 
present.  Introduced originally
 to explain the dynamics of galaxies, Dark Matter has found an established
 place in the Cosmological Model. Still, many questions and difficulties
 remain, see e.g.~\cite{Kro10}. In this context, also alternative explanations
  are discussed, one of them being MOND
 (modified-Newtonian-dynamics)~\cite{MOND1}.
MOND assumes, that the gravitational law is modified for small values of
the acceleration in the following way:
\begin{equation}
a_N=a \cdot \mu(a/a_0)
\end{equation}
Here, $a$ is the acceleration according to MOND,  $a_N$ is the
Newtonian acceleration $a_N=G\cdot m/r^2$, and $a_0=1.2 \cdot \rm{10^{-10} m
\, s^{-2}}$ is assumed to be a universal constant~\cite{Kro10},~\cite{Sca03}.
 The interpolation function $\mu(a/a_0)$ is
$\mu \to 1$ for $a>>a_0$, recovering Newton's Law, and $\mu \to a/a_0$ for
$a<<a_0$. Apart from these asymptotic values the interpolation function
is not determined by the theory, but has to be constrained by
 data. \\

A relativistic formulation incorporating the MOND theory has been
developed by Bekenstein~\cite{M5}. \\

MOND has so far passed many astronomical tests~\cite{Kro10}~\cite{Sca03}.
Apart from modifying Newton's Law, MOND could also be interpreted
as a violation of Newton's second axiom $F=m \cdot a$~\cite{MOND2},
 irrespective of
the nature of the force $F$. This latter aspect has been experimentally
checked, using electromagnetic restoring forces, and Newton's axiom verified 
down to accelerations of $3 \cdot 10^{-11} m \, s^{-2}$~\cite{Abr86} and
 $5 \cdot 10^{-14} m \, s^{-2}$~\cite{Gun07}. Therefore, a possible
 modification
according to MOND must rest with the gravitational force alone. This experiment
is designed to test Newton's Law at accelerations of the order $a_0$, using
only gravitational forces. It has
been argued, that such a test is not meaningful in the strong gravitational
field of the earth, but, due to a lack of a deeper understanding of MOND, this
view is not shared by everybody (see e.g.~\cite{Sca03}).  

\section{Experimental Procedure}

\begin{figure}[ht]%
\begin{center}
\includegraphics[width=10cm]{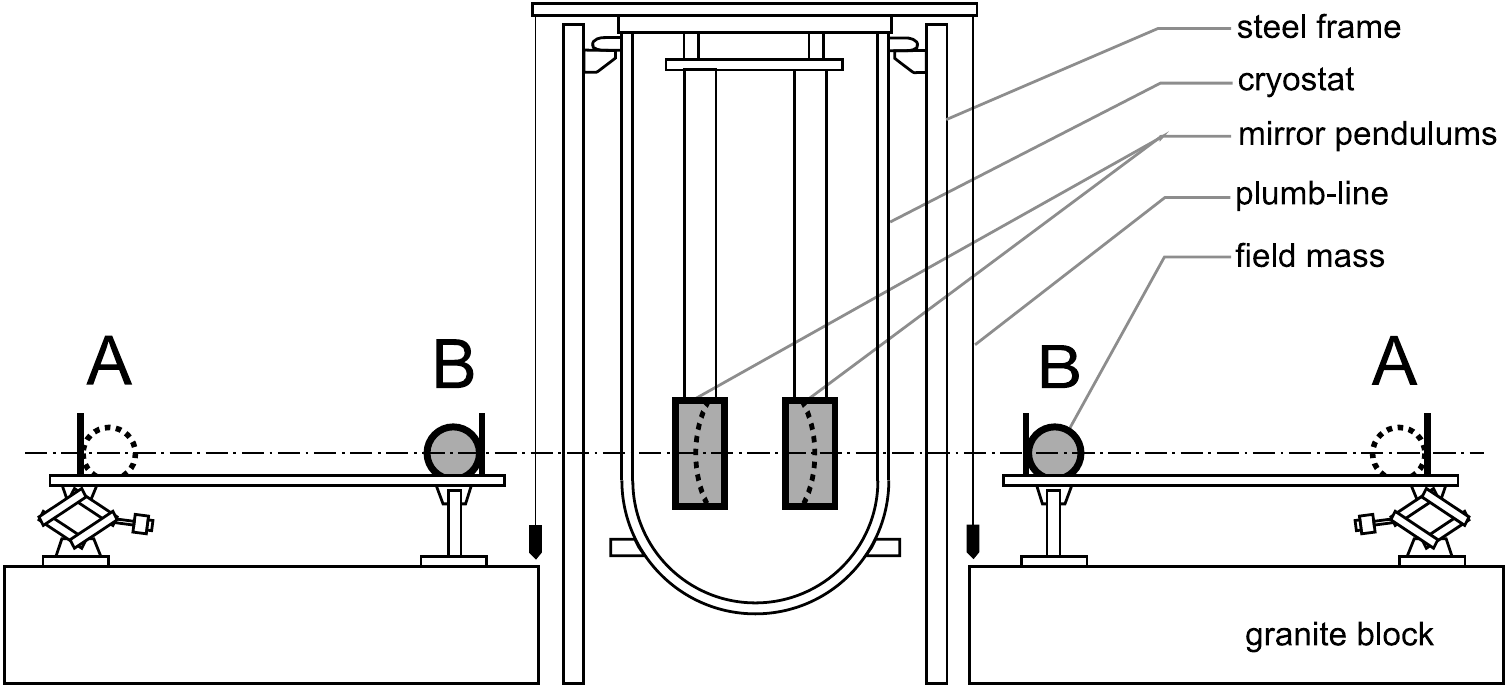}%
\caption[Experiment]{Schematic view of the experiment}%
\label{fig-exp1}%
\end{center}
\end{figure}

A schematic view of the experiment is shown in Fig.~\ref{fig-exp1}.

\begin{figure}[ht]%
\begin{center}
\includegraphics[width=10cm]{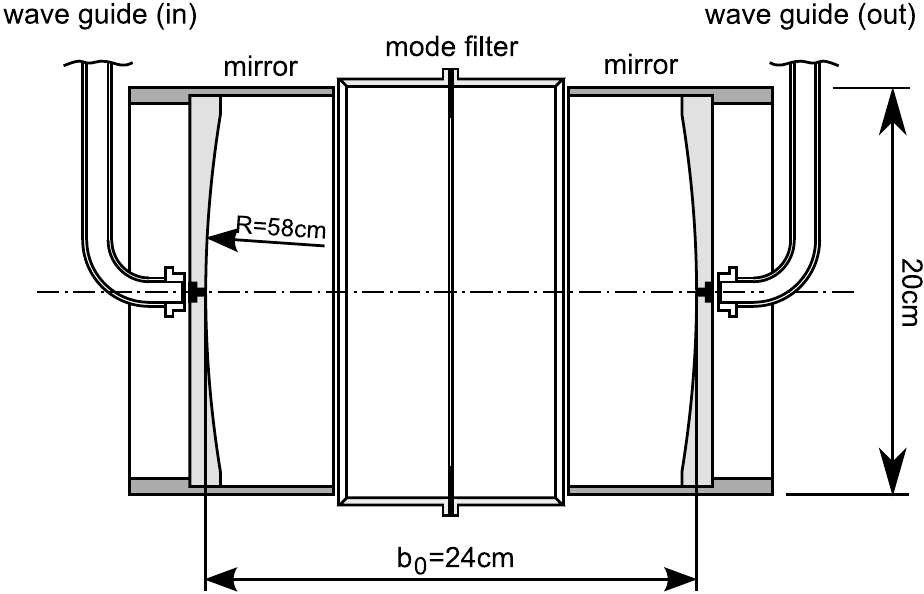}%
\caption[Experiment]{Schematic view of the resonator}%
\label{fig-exp2}%
\end{center}
\end{figure}

The central part of the experiment is a microwave resonator, tuned
to a frequency of about 21.3905 GHz. The resonator consists of two mirrors
with spherical surfaces, suspended by tungsten wires of about 3 m length,
resulting in a pendulum period of $3.289 \pm 0.010$  s.
  This part of the detector sits
in an evacuated vessel. Two field-masses are positioned outside the vacuum 
vessel on either side of the resonator, and are periodically and simultaneously
 moved between
a far (A) and a near (B) position. Their gravitational pull results in a
 small change of the position of the two mirrors, which is measured from the
 change of the resonance frequency.\\

 A detailed view of
the resonator is presented in 
Fig.~\ref{fig-exp2}, showing also the microwave 
guides. \\

The apparatus had been built and operated at Wuppertal University for
a precision measurement of the gravitational constant~\cite{Kle99},
\cite{Kle02},\cite{Wal95},\cite{Sch92},\cite{Asc99}. It was later tranferred
to DESY and reinstalled with some improvements for the 
stability of the support~\cite{Sch08},~\cite{Sch11}. \\

Measurements were carried out with three pairs of field-masses, consisting of 
spheres
of brass, marble and plastic, with masses of 9.02 kg, 2.92 kg and 1.00 kg,
respectively. All spheres have the same diameter of 12.7 cm. They were placed
at the near position with their centers at 76.6 cm on the left
 side and of 77.9 cm
on the right side from the center of gravity of the nearest mirror,
respectively. The acceleration of a mirror caused by the closer field-mass
 at the near position
 was $10.2 \cdot 10^{-10}m \, s^{-2}$, $3.3 \cdot
 10^{-10} m \,
 s^{-2}$, and $1.1 \cdot 10^{-10} m \, s^{-2}$ for the three masses, resulting
in a change of the distance between the two mirrors from about 0.210 nm
to  0.023 nm. If
Newton's Law is correct, the deflections due to the three field-masses
must be precisely proportional to their mass values. \\ 

Measurements were carried out by moving the left (right) field-masses from the
near position at 76.6 cm (77.9 cm) to a far position at 213 cm (220 cm) every 
40 min. \\

Measurements of the resonance frequency $f_R$ were performed every 2 sec
by tuning the frequency of the generator to five values around the resonance
 frequency and recording the resulting amplitude at the exit of the resonator.
 The resonance frequency was then determined by fitting a resonance curve of
 the form Equ.(2) to the five amplitude values $U(f)$.

\begin{equation}
U(f) =U_{max} \cdot \frac{1}{1 + 4 ((f-f_R)/f_w)^2} .
\end{equation}

Here U(f) is the amplitude at frequency f, $f_R$ is the fitted resonance
frequency and $f_w$ is the resonance width. \\

 The frequency measurements were then averaged over typically 1-2 min.
 The temperature at the apparatus was kept constant
 to about 0.1 degree; still the data show a strong drift with temperature,
which must be corrected for.

\begin{figure}[!htp]%
\begin{center}
\includegraphics[width=15cm]{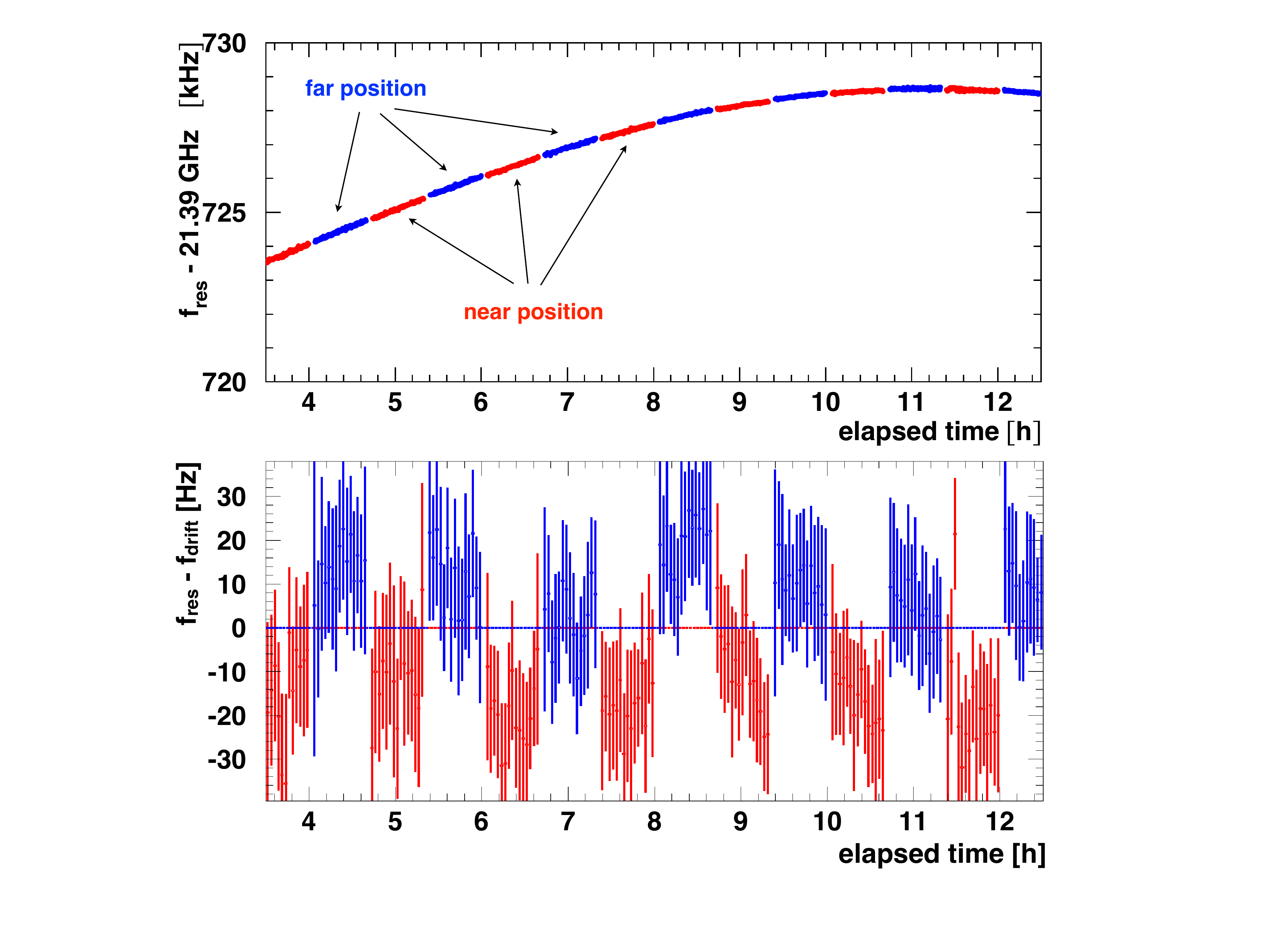}%
\caption[Experiment]{Frequency as a function of time for a measurement with the
 9.02 kg sphere, without (above) and
 after (below) a background drift subtraction. A large constant frequency
offset has been subtracted. }%
\label{fig-messg1}%
\end{center}
\end{figure}

Figure~\ref{fig-messg1} shows an example of a measurement with the 9.02 kg 
field-masses before and after
 subtraction of a slow frequency drift. From the frequency measurements a 
large constant frequency (about 21 GHz) has been subtracted.

\begin{figure}[!hbtp]%
\begin{center}
\includegraphics[width=12cm]{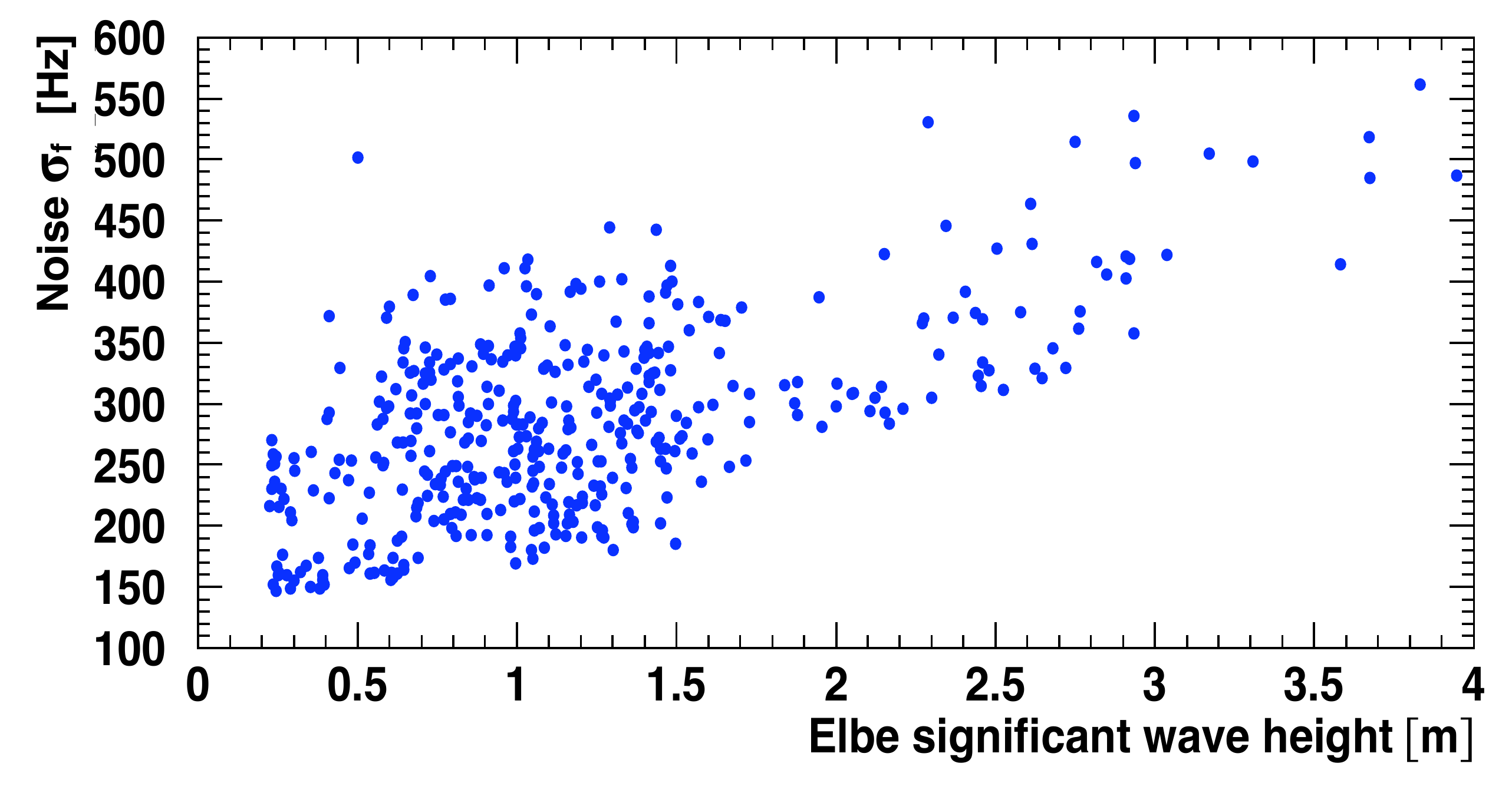}%
\caption[Experiment]{The noise (rms) of a single frequency measurement as a 
function of the significant wave height at the mouth of the Elbe river.}%
\label{fig-wave}%
\end{center}
\end{figure}

 Additional distortions come also from  sources
like ground movements, occasional earthquakes and waves from the North Sea: 
Fig.~\ref{fig-wave} shows as an example the rms noise of single frequency 
measurements
plotted against the significant wave height at the mouth of the Elbe
 river.

\section{Results}

The data were evaluated using six different methods of background
subtraction, to deal with slow drifts of the resonance frequency and with
short-term background variations: Overall polynomial fits (A1, A3), piecewise
 3rd order polynomial
fit (A5), piecewise 5th order polynomial fit (A6), and sliding average 
(A2, A4)~\cite{Sch11}.
For each mass the uncertainties of the mean values in Table 1 were determined
 from the variance of the results of about 25
independent data runs with an average duration of 12 hours. The differences
 between the results of different
methods for the 9.02 kg data are somewhat larger than expected from the 
individual errors and reflect 
differences in dealing with short-term variations of the
 background. To deal with it, the uncertainty of the 9.02 kg data is in the
following Fig.~\ref{fig-mond} and in the calculation of G increased by a
factor of 1.5, according to the PDG prescription~\cite{PDG}. 
 All six methods were checked with several sets of Monte Carlo data, to which
regular and irregular noise similar to the one observed in the data had been
 added. All methods were able to reproduce the correct input signal within
the uncertainties~\cite{Sch11}. \\ 

 Method A1 was used as the central value and the
other methods as consistency checks. \\

\begin{table}[h]%
\centering
\caption{ Mean frequency shift $\Delta f$ in Hz for the three field-masses}
\begin{tabular}{|l|l|l|l|}
\hline
Method   & $\Delta f$ 1.0 kg &$\Delta f$ 2.92 kg & $\Delta f$ 9.02 kg \\
\hline
A1       & 2.06 $\pm$ 0.25   & 6.09 $\pm$0.24    & 18.65 $\pm$ 0.40 \\
A2       & 2.00 $\pm$ 0.44   & 6.25 $\pm$0.32    & 19.12 $\pm$ 0.54 \\
A3       & 2.06 $\pm$ 0.38   & 6.19 $\pm$0.34    & 18.53 $\pm$ 0.60 \\
A4       & 2.94 $\pm$ 0.48   & 6.36 $\pm$0.31    & 19.30 $\pm$ 0.61 \\
A5       & 1.72 $\pm$ 0.59   & 6.31 $\pm$0.36    & 20.09 $\pm$ 0.51 \\
A6       & 1.60 $\pm$ 0.25   & 5.63 $\pm$0.23    & 17.00 $\pm$ 0.56 \\
\hline
\end{tabular}
\end{table}

Assuming, that the potential effect due to MOND is negligeable for
the 9.02 kg field-masses, the gravitational constant G can be computed from the
frequency shift $\Delta f$ between the far and near positions of the 
9.02 kg field-masses. The values of $\Delta f$ given in Table 1 are the sum of
 the frequency shifts of the right and left field-masses.

The frequency shift $\delta f$ due to one field-mass is given by

\begin{equation}
\frac{\delta f}{f}=\frac{GMT_0^2\cdot \Delta(1/r^2)}{4 \pi^2 b} 
\end{equation}

with

\begin{equation}
\Delta(1/r^2)= (1/r_n^2 -1/(r_n+b)^2)-(1/r_f^2-1/(r_f+b)^2)
\end{equation}

Here $f$ is the frequency, $M$ the field-mass, $T_0$ the pendulum period of
the cavity, $b$ the distance between the two cavities, $r_n$ and $r_f$ are
the distances of the near and far position of the field-mass from the nearest 
mirror, respectively.

Using method A1 as a reference value, and after a correction taking account
of the detailed shape of the mirrors~\cite{Sch11}, one obtains a value for
 $G$  \\

$G$ = (6.57 $\pm 0.21$ $\pm 0.11) \cdot 10^{-11}m^3/kg\, s^2$ \\

where the first uncertainty is due to the uncertainty of the frequency shift.
The second uncertainty is systematic, with the list of systematic uncertainties
 as given in the table below. The first three entries in the table follow 
directly from the
corresponding measurement uncertainties, the fourth entry follows from the 
estimated accuracy of the integration over the mirrors, and the uncertainty
of $b$ was determined from an analysis of the mode spectrum of the resonator. 
\\

\begin{table}[h]%
\centering
\caption{Systematic uncertainties contributing to the measurement of $G$.}
\begin{tabular}{|l|l|}
\hline
Source   & uncertainty in \% \\
\hline
pendulum frequency $T_0^{-1}$                         & 0.66 \\
position of the field-masses $r_n,r_f$                & 1.48  \\
value of the field-masses $M$                         & 0.11  \\ 
integration over the mass distribution of the mirrors & 0.20  \\
distance between mirrors $b$                          & 0.01  \\
\hline
\end{tabular}
\end{table}

This value of $G$ agrees  with the world average~\cite{PDG} of
$G=6.67428(67)\cdot 10^{-11} m^3/kg\, s^2$ within the uncertainties. \\ 

Predictions from  MOND are not unambiguous. We assume, that the forces
from each field-mass on each cavity, as calculated from the MOND formula,
can be added linearly. With this assumption one can calculate the frequency
shifts for the different field-masses and for the different interpolation
functions $\mu(x)$, 
with $a$=acceleration due to MOND, $a_N$=Newton's acceleration, $x=a/a_0$,
 $y=\sqrt{(a_N/a_0)}$:  \\

\begin{displaymath}
\begin{array}{lrl}
MOND1~\cite{MOND2}  &  \mu(x)=  &  x/\sqrt{1+x^2}  \\ 

MOND2~\cite{M2} & \mu(x)=  &  x/(1+x)   \\ 

MOND3~\cite{M3}  &  \mu(x)=  &  6x/\pi^2 \cdot\int^{\pi^2/6x}_0
 z/(e^z-1) dz   \\

 MOND4~\cite{M4} &  a=  &  a_0 y \cdot (1-y^4)/(1-y^3)  \\

 MOND5~\cite{M5} &  \mu(x)=  &  (\sqrt{1+4x}-1)/(\sqrt{1+4x}+1)  \\

\end{array}
\end{displaymath}

\vspace{0.5cm}
Fig.~\ref{fig-mond} shows a comparison of the measurements with the predictions
of the five interpolation functions.
 The normalized frequency shift
$\Delta f/M$ (where $M$ is the mass of the field-mass) is plotted for the three
 field-masses. The error bars on the data points represent the uncertainty of 
the measured frequency shifts.
 The width of the bands for the
 interpolating functions shows the effect of the systematic uncertainties.
 If Newton's Law is valid, $\Delta f/M$ must be the same for all
field-masses.
 Version MOND5 due to Bekenstein's relativistic theory is clearly
 ruled out (also ruled out by astrophysical observations~\cite{M2}).
The versions MOND1 and MOND2 are only slightly disfavoured. \\

%\newpage

\begin{figure}[!hbtp]%
\begin{center}
\includegraphics[width=12cm]{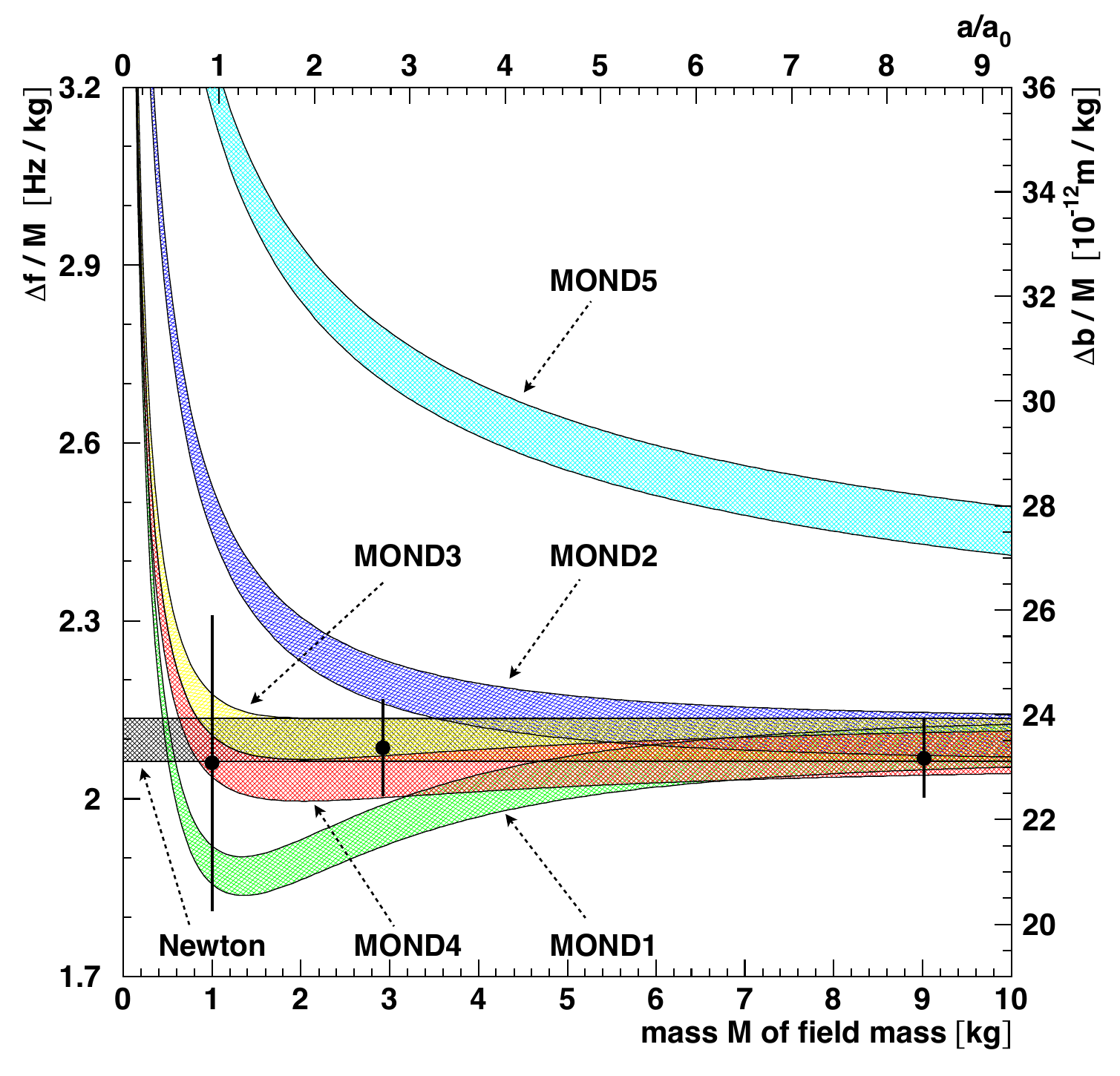}%
\caption[Experiment]{Comparison of different versions of the 
MOND interpolation function with the measurements. The values of the 
field-masses $M$ are plotted against $\Delta f/M$, where $\Delta f$ is
the frequency shift. The upper horizontal axis indicates the value of the 
acceleration of one of the mirrors caused by the closer field-mass in the
near position in units of the MOND acceleration $a_0$. The vertical axis on 
the right hand side shows the corresponding relative change of the distance
between the mirrors.}
\label{fig-mond}%
\end{center}
\end{figure}

%\newpage

\section{Conclusions.}
Newton's Law of Gravitation has been tested for small values of the
acceleration, using a pair of pendulums to measure the gravitational 
attraction.
The data cannot refute the MOND theory, but they have successfully probed
Newton's Law down to the MOND acceleration $a_0 = 1.2 \cdot 10^{-10}\, m/s^2$.
The accuracy of the measurements will be improved by moving the 
experiment to an underground location and by a better mechanical support. \\ 

%\newpage

\large{\bf Acknowledgments} \\
\normalsize

It is a pleasure to thank the DESY Directorate for their support. We
also acknowledge the assistance of U.Cornett, M.Gil, D.Habercorn,
 T.K\"ulper, C.Muhl, J.Schaffran, H.J.Seidel and V.Sturm in setting up the
experiment, and we thank the technical groups of DESY for their help and advice
 in many technical questions.

\end{document}